\documentstyle[12pt]{article}
\begin{document}
\thispagestyle{empty}

\newcommand{\be}{\begin{equation}}
\newcommand{\ee}{\end{equation}}
\newcommand{\sect}[1]{\setcounter{equation}{0}\section{#1}}
\renewcommand{\theequation}{\thesection.\arabic{equation}}
\newcommand{\vs}[1]{\rule[- #1 mm]{0mm}{#1 mm}}
\newcommand{\hs}[1]{\hspace{#1mm}}
\newcommand{\mb}[1]{\hs{5}\mbox{#1}\hs{5}}
\newcommand{\bea}{\begin{eqnarray}}
\newcommand{\ena}{\end{eqnarray}}

\newcommand{\wt}[1]{\widetilde{#1}}
\newcommand{\und}[1]{\underline{#1}}
\newcommand{\ov}[1]{\overline{#1}}
\newcommand{\sm}[2]{\frac{\mbox{\footnotesize #1}\vs{-2}}
		   {\vs{-2}\mbox{\footnotesize #2}}}
\newcommand{\prt}{\partial}
\newcommand{\eps}{\epsilon}

\newcommand{\R}{\mbox{\rule{0.2mm}{2.8mm}\hspace{-1.5mm} R}}
\newcommand{\Z}{Z\hspace{-2mm}Z}

\newcommand{\cd}{{\cal D}}
\newcommand{\cg}{{\cal G}}
\newcommand{\ck}{{\cal K}}
\newcommand{\cw}{{\cal W}}

\newcommand{\vj}{\vec{J}}
\newcommand{\vl}{\vec{\lambda}}
\newcommand{\vz}{\vec{\sigma}}
\newcommand{\vt}{\vec{\tau}}
\newcommand{\vw}{\vec{W}}
\newcommand{\poiss}{\stackrel{\otimes}{,}}


\newcommand{\NP}[1]{Nucl.\ Phys.\ {\bf #1}}
\newcommand{\PL}[1]{Phys.\ Lett.\ {\bf #1}}
\newcommand{\NC}[1]{Nuovo Cimento {\bf #1}}
\newcommand{\CMP}[1]{Comm.\ Math.\ Phys.\ {\bf #1}}
\newcommand{\PR}[1]{Phys.\ Rev.\ {\bf #1}}
\newcommand{\PRL}[1]{Phys.\ Rev.\ Lett.\ {\bf #1}}
\newcommand{\MPL}[1]{Mod.\ Phys.\ Lett.\ {\bf #1}}
\newcommand{\BLMS}[1]{Bull.\ London Math.\ Soc.\ {\bf #1}}
\newcommand{\IJMP}[1]{Int.\ Jour.\ of\ Mod.\ Phys.\ {\bf #1}}
\newcommand{\JMP}[1]{Jour.\ of\ Math.\ Phys.\ {\bf #1}}
\newcommand{\LMP}[1]{Lett.\ in\ Math.\ Phys.\ {\bf #1}}

\renewcommand{\thefootnote}{\fnsymbol{footnote}}

\newpage
\setcounter{page}{0}
\pagestyle{empty}

\vs{30}

\begin{center}

{\LARGE {\bf Generalized NLS Hierarchies}}\\
[.3cm]

{\LARGE {\bf from Rational ${\cw}$ Algebras}}\\[1cm]

\vs{10}

\vs{10}

{\large {F. Toppan}}\\
\quad \\
{\em Laboratoire de Physique Th\'eorique }\\
{\small E}N{\large S}{\Large L}{\large A}P{\small P}\\
{\em ENS Lyon, 46 all\'ee d'Italie,} \\
{\em F-69364 Lyon Cedex 07, France.}

\end{center}
\vs{20}

\centerline{ {\bf Abstract}}

\indent

Finite rational $\cw$ algebras are very natural structures appearing
in coset constructions when a Kac-Moody subalgebra is factored out.
In this letter we
address the problem of relating these algebras to integrable
hierarchies of equations, by showing how to associate to a rational
$\cw$ algebra its corresponding hierarchy. We work out two examples:
the $sl(2)/U(1)$ coset,
leading to the Non-Linear Schr\"{o}dinger hierarchy, and the $U(1)$
coset of the Polyakov-Bershadsky $\cw$ algebra, leading to a
$3$-field representation of the
KP hierarchy already encountered in the literature. In such examples
a rational algebra appears as
algebra of constraints when reducing a KP hierarchy to a finite field
representation. This fact arises the natural question whether
rational algebras
are always associated to such reductions and whether a classification
of rational algebras can lead to a classification of the integrable
hierarchies.

\vfill
\rightline{{\small E}N{\large S}{\Large L}{\large A}P{\small
P}-L-448/93}
\rightline{November 1993}

\newpage
\pagestyle{plain}
\renewcommand{\thefootnote}{\arabic{footnote}}
\setcounter{footnote}{0}

\sect{Introduction}

\indent

In a previous work \cite{DFSTR} it has been shown that a natural
structure
of finite rational $\cw_Q $ algebra (the index $Q$ stands for
``quotient") appears when looking to the subsector in the
enveloping algebra
of an affine Lie $\cal G$ or standard polynomial $\cw$ algebra which
commutes with respect to
a given Kac-Moody subalgebra ${\hat{\cal G}}\subset {\cal G}, \cw$.
Such rational $\cw_Q$ algebras can also be seen as specific
realizations of non-linear $W_\infty$ algebras of polynomial type.

In this letter we will point out that to each such rational
$\cw_Q$ algebra is associated a whole hierarchy of equations
admitting an infinite
number of hamiltonians in involution. These hierarchies could deserve
the name of generalized Non-Linear Schr\"{o}dinger (NLS) hierarchies.
The simplest
example of our framework is indeed provided by the Non-Linear
Schr\"{o}dinger hierarchy, which is obtained from the coset
construction
$sl(2)/ U(1) $. Therefore we will refer to these generalized
hierarchies also as coset hierarchies and to their associated finite
$\cw_Q$ algebras as rational
coset $\cw$ algebras\footnote{For rational $\cw$ algebras see also
\cite{feher}.}.

Some comments are in order: the arising of a $\cw_\infty$ algebra
structure
in the Witten's black hole \cite{witten} $sl(2)/U(1)$ coset model has
been pointed out in many papers \cite{{bakas},{yuwu}} (for another
coset construction leading to $\cw_\infty$ algebra see
\cite{ivanov}). In \cite{yuwu} a realization
in terms of parafermions has been given. For a connection of such
construction to the NLS hierarchy see e.g. \cite{schiff}. \\
The fact that an algebra with infinite elements is produced out of a
finite number of parafermions is already an indication of the
``degeneracy" of such algebra. Moreover, $\cw_\infty$-algebra
structures appear as algebras of constraints when reducing
KP hierarchies to standard KdV-like hierarchies (see e.g.
\cite{{aratyn},{bonora},{bonora2}}): in this framework higher order
constraints are just a consequence of a finite set
of constraints. Here again the ``degenerated" character of such
$\cw_\infty$ algebras appears. Rational $\cw $ algebras are nothing
else than $\cw_\infty$ algebras which can be explicitly ``solved",
namely that can be expressed in closed form with a finite number of
fields and a finite number of algebraic constraints among them which
are consistent with the Poisson brackets structure of the underlining
$\cw_\infty$ algebra.
In \cite{DFSTR} a systematic way to produce rational algebras has
been discussed. Here we will rely on the results of \cite{DFSTR} to
provide a framework for generalizing the works of
\cite{{yuwu},{schiff}}.

To be specific we will treat only cosets obtained by quotienting out
an
abelian $U(1)$ Kac Moody subalgebra, but the extension to non-abelian
quotients
is plainly straightforward. The whole discussion will be done for the
classical
case, so that the algebras involved are assumed to be Poisson
brackets algebras.
Results and conceptual points concerning quantum rational $\cw$
algebras will be reported elsewhere.

A technical remark: we are interested in Poisson brackets of
invariant quantities (with respect to the quotient subalgebra)
defined in a manifestly covariant way. For that reason our framework
is particularly suitable for computations because we do not need to
take into account Dirac's brackets
but only the original Poisson brackets structure.

It should be noticed that, despite striking analogies, namely
two-bosons realization of the hierarchies, introduction of covariant
derivatives, etc., our construction is basically different from the
one proposed in \cite{aratyn}: the latter is based on currents
realized through bosonic $\beta-\gamma$ systems,
while our currents are standard Kac-Moody currents.

The plan of this paper is the following: in section $2$ the basic
features of
rational coset $\cw_Q$ algebras are recalled, the connection to
non-linear $W_\infty$ algebras is established, the formalism and
conventions needed for later use are introduced. In section $3$ the
NLS hierarchy is revisited as a realization of
the simplest possible example of $\cw_Q$ algebra, based on the
$sl(2)/U(1)$ coset; this is intended to illustrate  the method which
can be straightforwardly
generalized to produce other coset hierarchies. In section $4$ as an
application
of the above method, we will discuss the hierarchy associated to the
simplest coset obtainible from a polynomial $\cw$
algebra: this is the $U(1)$ coset of the Polyakov-Bershadsky
\cite{polya}
$\cw$ algebra, which is of type $(1,{\textstyle{3\over 2}}
,{\textstyle{3\over 2}}, 2)$. The hierarchy associated to such
rational $\cw$ algebra turns out to coincide with the first reduction
of the four-field KP hierarchy of ref.\cite{bonora}.

\sect{Rational $\cw$ algebras}

\indent

In this section we will review the formulation of rational $\cw_Q$
algebras
as stated in \cite{DFSTR}. For simplicity we will consider only
abelian ($U(1)$)
quotients.

Let ${\cal G}$ or $\cw$ denote respectively a Kac-Moody or a $\cw$
algebra
admitting a subalgebra generated by a  $U(1)$ Kac-Moody current
$J(z)$:
\be
\{J(z), J(w)\} = \gamma \delta'(z-w)\equiv \gamma\prt_w\delta(z-w)
\ee
(in the classical case the normalization factor $\gamma$ can be fixed
without loss of generality as will be done in the following).
It is therefore possible to express any other element of the algebra
in terms of a basis of fields $W_{i,q_i}$ having a definite charge
$q_i$ with respect to $J(z)$, namely satisfying:
\be
\{J(z), W_{i,q_i}(w)\} = q_i W_{i,q_i}\delta(z-w)
\ee
A derivative $\cd$, covariant with respect to the above relations,
can be introduced:
\be
\cd W_{i,q_i}(w) = (\prt -{\textstyle{q_i\over \gamma}} J(w))
W_{i,q_i}(w)
\ee
The elements in $Com_J({\cal G},\cw)$, the subalgebra of the
enveloping algebra  commuting with $J(w)$, \footnote{ It is better
conceptually to understand the derivative operator $\prt
={\textstyle{d\over dw}}$ as an element of the original algebra, as
this is the case for Kac-Moody algebras.} are therefore spanned by
the vanishing total charge monomials \\
$(\cd^{n_1}V_{1,q_1})(\cd^{n_2}V_{2,q_2})...
(\cd^{n_j}V_{j,q_j})$, where the $n_i$'s are non negative integers
and the total
charge is $q=q_1+q_2+...q_j =0$.\\
{}From now on we will concentrate only on invariants produced by
bilinear combinations such as
\begin{eqnarray}
\cd^pV_+\cdot\cd^qV_-&&
\label{bilin}
\end{eqnarray}
(with $p,q\geq 0$ and $V_\pm$ have opposite charges),
together with of course originally invariant fields.\footnote{This is
only in order to avoid unnecessary technical complications in the
following discussion, but hierarchies associated to multilinear
invariants can be produced along the same lines as for the bilinear
ones and are worth being studied.} \\
We summarize here the basic results of \cite{DFSTR}, with some extra
comments:\\
{\it i}) there exists a linear basis of fields, given by $V^{(p)}=
\cd^p V_+\cdot V_- $ such that any bilinear invariants of the kind
(\ref{bilin}) is a linear combination
of the $V^{(p)}$'s and the derivatives acting on them.\\
{\it ii}) the Poisson brackets algebra of the fields $V^{(p)}$'s
among themselves is
closed (possibly with the addition of other invariants, in the
general case),
but never in a finite way (the Poisson brackets of $V^{(p)}$ with $
V^{(q)}$ necessarily generates on the right hand side terms depending
on $V^{(p')}$, with $p'> p,q $). Moreover it can be explicitly
checked
that it is a non-linear algebra, so that it has the structure of a
non-linear $W_\infty$
algebra.\\
{\it iii}) due to the properties of the covariant derivative, the
fields $V^{(p)}$, which are linearly independent, satisfy algebraic
relations like the following quadratic ones
\bea
V^{(p+1)} \cdot V^{(0)} &=& V^{(0)} \cdot \partial V^{(p)} - V^{(p)}
\cdot
\partial V^{(0)} + V^{(p)} \cdot V^{(1)}
\label{rela}
\ena
Such relations allow to express algebraically the fields $V^{(p)}$,
for $p\geq 2$ in terms of the fundamental fields $V^{(0)}$ and
$V^{(1)}$. The above derived non-linear $\cw_\infty$ algebra has
therefore the structure of a rational $\cw $ algebra. Notice that
relations like (\ref{rela}) contain no informations if the fields
$V_\pm$ are fermionics. (\ref{rela}) can still be applied to
superalgebras if $V_\pm$ are bosonic superfields.
\\
{\it iv}) if $Com_J({\cal G}, \cw)$ contains a field $T(w)$ (the
stress-energy tensor)
whose Poisson brackets are the Virasoro algebra with non-vanishing
central charge, and moreover $V^{(0)}$ is primary with conformal
dimension $h$, then there exists a one-to-one correspondence between
the fields $V^{(p)}$ of the basis, and an infinite tower of uniquely
determined fields $W_{h+p}$, primary with respect to $T$, with
conformal dimension $h+p$. This relation should be understood as
follows: $V^{(p)}$ is the leading term in the associated primary
field. The remaining terms are fixed without ambiguity, some of
them just requiring $W_{h+p}$ being primary, some others once
a specific scheme to determine them is adopted (as an analogy, one
should think to the choice of the renormalization scheme when dealing
with renormalizable quantum field theories).\\
 As we will see, the condition of having a
non-vanishing central charge drops for the $sl(2)/U(1)$ coset model:
therefore no infinite tower of primary fields associated to
each $V^{(p)}$ can be generated
(we have an infinite tower of ``almost" primary fields
associated to them). An infinite number of primary fields can still be
produced, but they are of a trivial type, being just products of lower
order primary fields.
 Anyway the
structure of
a rational algebra with two primary fields is mantained even in
this case.  The next
simplest model admitting an infinite tower of invariant primary
fields associated to $V^{(p)}$ is based on the coset
${\textstyle {sl(2)\times U(1)\over U(1)}}$, since now there
exists another $U(1)$ current,commuting with $J(w)$, which allows to
define an invariant stress-energy tensor with non-vanishing charge.

In the following, in order to illustrate how our procedure works we
will treat explicitly two examples, so let us write down their
algebra here.

\subsection{ case a: the ${\textstyle{sl(2)\over U(1)}}$ coset.}

The classical $sl(2)-{\cal KM}$ algebra is given by the following
Poisson brackets:
\bea
\{J_+(z),J_-(w)\} &=&  \delta'(z-w) - 2 J_0 (w) \delta(z-w) \equiv
\cd (w)\delta(z-w) \nonumber \\
 \{J_0(z), J_\pm (w)\} &=& \pm J_\pm (w) \delta (z-w)
 \nonumber \\
 \{J_0 (z),J_0(w)\} &=& {-\textstyle{1\over
 2}}\delta'(z-w)\nonumber\\
\{J_\pm(z),J_\pm(w)\} &=& 0
\label{KM}
\ena
 $J_\pm$ play the role here of the fields $V_\pm$ in the previous
 subsection; they have conformal dimension $1$
(here and in the following, the symbol $\delta'(z-w)$ is understood
as $\prt_w\delta (z-w)$).\\
The rational coset algebra of the commutant with respect to the $J_0$
current is given by the following Poisson brackets
\bea
\{W_2(z),W_2(w)\} &=& 2 W_2(w) \delta'(z-w) + \prt W_2(w) \delta(z-w)
\nonumber \\
\{W_2(z),W_{3}(w)\} &=& 3 W_{3}(w) \delta'(z-w) + \prt
W_{3}(w) \delta(z-w) \quad \nonumber \\
\{W_3(z),W_3(w)\} &=& 2 W_2(w) \delta'''(z-w) + 3 \prt W_2(w)
\delta(z-w)'' +\nonumber\\
&& [16 V^{(2)} - 8 \prt W_3 +8 {W_2}^2 -3 \prt^2 W_2 ] (w)
\delta'(z-w)+
\nonumber \\
&&\prt_w [8 V^{(2)} - 4 \prt W_3 +4 {W_2}^2 -2\prt^2 W_2 ] (w)
\delta(z-w)\nonumber\\
&&
\label{coset}
\ena
We have preferred to express the above algebra in the basis of
(uniquely determined) primary fields
$W_2 = J_+\cdot J_-$ and $W_3 = \cd J_+ \cdot J_- - J_+ \cdot \cd J_-
$.
They have dimension $2,3$ respectively, while
\bea
V^{(2)} &=& \cd^2 J_+ \cdot J_- = {1\over 4 W_2}[ W_3^2 +2 W_2\prt
W_3 + 2 W_2\prt^2W_2 -\prt W_2\prt W_2].
\ena
The second equality follows from the relation (\ref{rela}).\\
$W_2$ plays the role of a stress-energy tensor having no central
charge.
As already stated, in this simple example there exists no infinite
tower of primary fields associated to
the $V^{(p)}$'s fields, the only primary ones being $W_{2,3}$
and their products ${W_2}^m {W_3}^n$ for $m,n$ non-negative
integers.

The algebra of the fields $V^{(p)} = \cd^p J_+\cdot J_-$ is a
non-linear $\cw_\infty$ algebra: if we let from the very beginning
identify $J_0\equiv 0$,
then $J_\pm$ can be identified with the fields $\prt \beta$ and
$\gamma$ of a bosonic $\beta -\gamma$ system, the covariant
derivative in $V^{(p)}$ must be replaced by the ordinary derivative
and the non-linear $\cw_\infty$ algebra is reduced to the standard
linear $w_\infty$ algebra.

\subsection{case b: The $U(1) $ Polyakov-Bershadsky coset.}

In this subsection we introduce the $U(1)$ commutant of the
Polyakov-Bershadsky $\cw$ algebra\cite{polya}. This algebra is
associated to a non-abelian $sl(3)$ Toda model. For constructing
general $\cw$ algebras from Toda theories, both abelian and
non-abelian, see e.g. \cite{gervais,sorba}.

The Polyakov-Bershadsky $\cw$ algebra is defined in terms of a
stress-energy tensor $T(w)$, a $U(1)$ Kac-Moody current $J(w)$ and
two bosonic charged fields
of (covariantly conformal) dimension ${\textstyle{3\over 2}}$. It is
explicitly given by the
following Poisson brackets:
\bea
\{J(z),J(w)\} &=&  {\textstyle{3\over 2}} \delta'(z-w) \nonumber \\
\{T(z),T(w)\} &=& 2 T(w) \delta'(z-w) + \prt T(w) \delta(z-w) -
{\textstyle{1\over 2}}\delta'''(z-w) \nonumber \\
\{T(z),J(w)\} &=& 0 \nonumber \\
 \{T(z),W_{\pm}(w)\} &=& {\textstyle{3\over 2}} W_{\pm}(w)
 \delta'(z-w) + (\cd
W_{\pm})(w) \delta(z-w) \nonumber \\
 \{J(z),W_{\pm}(w)\} &=& \pm {\textstyle{3\over 2}} W_{\pm}(w)
 \delta(z-w)
\nonumber \\
 \{W_+(z),W_-(w)\} &=& (T - \cd^2)(w) \delta(z-w)
\nonumber \\
 \{W_+(z),W_+(w)\} &=& \{W_-(z),W_-(w)\} = 0
\nonumber \\
&&
\ena
The covariant derivative is now $\cd W_\pm = (\partial \mp J) W_\pm
$.

The rational coset $\cw$ algebra of the $U(1)$ commutant of the above
algebra
has been written down in \cite{DFSTR}, the expression being given in
terms of primary fields.
Since, as already remarked, for the purpose of integrable hierarchies
is not necessary to dispose of a basis of primary fields, here we
prefer to express the coset algebra in terms of the fields $T(w)$ and
$V^{(n)} = \cd^n W_+\cdot W_-$, this linear basis being more suitable
for making computations (deriving equations of motions and so on).
The algebra
can be expressed in a closed form as follows
\bea
\{T(z),T(w)\} &=& 2 T(w) \delta'(z-w) + \prt T(w) \delta(z-w) -
{\textstyle{1\over 2}}\delta'''(z-w)
\nonumber \\
\{T(z),V^{(0)}(w)\} &=& 3 V^{(0)}(w) \delta'(z-w) + \prt
V^{(0)}(w) \delta(z-w) \quad
\nonumber \\
\{T(z), V^{(1)}(w)\} &=& {\textstyle{3\over 2}}
V^{(0)}(w)\delta''(z-w) +V^{(1)}(w) \delta'(z-w) +\prt
V^{(1)}(w)\delta(z-w)
 \nonumber \\
\{V^{(0)}(z),V^{(0)}(w)\} &=& (4V^{(1)}-2\prt V^{(0)})(w)\delta'(z-w)
+
\prt(2 V^{(1)} -\prt V^{(0)})(w)\delta (z-w)
\nonumber \\
\{V^{(0)}(z),V^{(1)}(w)\} &=& V^{(0)}(w) \delta'''(z-w) +2 V^{(1)}
\delta''(z-w) +
\nonumber\\
&& (V^{(2)}-2\prt V^{(1)})(w)\delta'(z-w) +
\nonumber \\
&& (2\prt V^{(2)} -\prt^2 V^{(1)} - V^{(0)}\prt T ) (w) \delta(z-w)
\nonumber\\
\{V^{(1)}(z),V^{(1)}(w)\} &=& 2 V^{(1)}(w) \delta'''(z-w) +
3\prt V^{(1)}(w)\delta''(z-w) +\nonumber\\
&& ( 3 V^{(3)} - 6\prt V^{(2)} + 3\prt^2 V^{(1)} -{\textstyle {3\over
2}} V^{(0)} V^{(0)} - 2 T V^{(1)}) (w) \delta'(z-w) +\nonumber\\
&& \prt( \prt^2 V^{(1)} - 3 \prt V^{(2)} + 3  V^{(3)} - TV^{(1)}
-{\textstyle{3\over 4}} V^{(0)} V^{(0)} ) (w) \delta (z-w)\nonumber\\
&&
\label{commalgebra}
\ena
The fields $T(w)$, $V^{(0,1)}$ are the fundamental ones. The fields
$V^{(n)}$
for $n\geq 2 $ being determined from (\ref{rela}).\\
Since in this case the invariant stress-energy tensor $T(w)$ admits a
non-vanishing central charge, the commutant contains an infinite
tower of primary fields and the rational algebra has an underlining
structure of a
non-linear $\cw_\infty$ algebra of {\it primary} fields.

\section{The NLS-hierarchy revisited.}

\indent

In this section we will derive the hierarchy associated to the
non-linear Schr\"{o}dinger equation
directly from the rational $\cw$ algebra (\ref{coset}).

Before doing that, a few words need to be spent: one very remarkable
feature of the rational algebras (\ref{coset}) and
(\ref{commalgebra})
is that, in both cases, it is possible to find out a basis of
generating fields for the rational algebra such that for any Poisson
brackets involving two fields of the basis, the term
in the right hand side proportional to the delta-function turns out
to be a total derivative. This property is of course immediately seen
for the rational
algebra (\ref{coset}) in the basis we have expressed it, but it is
valid also for (\ref{commalgebra}) as we will discuss in the next
section.
This very important property is shared by standard $\cw$ algebras
\cite{pope}. We will see that it is of crucial importance
for making a connection with the hierarchies and is not
unexpected.\footnote{as for the rational algebra (\ref{commalgebra}),
it is related to the fact that the Polyakov-Bershadsky algebra is
associated to a fractional KdV hierarchy, see
\cite{frac}.}
Even if we do not have at present a general argument stating that
this is always the case, it is indeed true that in any example of
coset construction worked so far the above property is satisfied.
Such examples include cosets realized from Kac-Moody and $\cw$
algebras, with respect to abelian and non-abelian Kac-Moody quotient,
cosets of superKac-Moody and super$\cw$ algebras, even cosets
realized for quantum
algebras. The discussion which follows, done for the
(\ref{coset})
algebra, can be generalized for any rational algebra satisfying the
above property, which means at least a very large class of rational
algebras.

Let us come back now to the rational algebra (\ref{coset}). Since we
are not
interested in its conformal property, it is more convenient to look
at it as
expressed in terms of the fields $V^{(n)}= \cd^n J_+ \cdot J_-$, for
$n$ non-negative integer. \\

The above mentioned property tells us that the line integrals $H_1
=\int dw V^{(0)}(w)$, $H_2 = \int dw V^{(1)}(w) $ have vanishing
Poisson brackets among themselves and can therefore be regarded as
two compatible
hamiltonians. \\
The equations of motion for the composite fields $V^{(n)}$ in the
commutant are equivalent to the equations of motion for the original
fields $J_{\pm,0}$. In one direction this statement is obvious; the
converse is also true: at first let us remark that,
by construction, $J_0$ commutes with any field in the commutant,
therefore its
equations of motion are always ${\textstyle {d\over d t}} J_0 =0$,
and is consistent
to set $J_0 =0$. Next, the following relations hold:
\begin{eqnarray}
{{\prt J_+\over J_+}} ={ {V^{(1)}\over V^{(0)}}};
\qquad && \quad {{\prt J_-\over J_-}} ={{\prt V^{(0)} -V^{(1)}\over
V^{(0)}}}
\end{eqnarray}
they tell that the dynamics of $J_\pm$ can be reconstructed from the
dynamics of
$V^{(0,1)}$ through the non-local transformations
\begin{eqnarray}
J_+(z) &=& e^{\int^zdw({\textstyle {V^{(1)}\over V^{(0)}}
})(w)}\nonumber\\
J_-(z) &=& e^{\int^z dw({\textstyle {\prt V^{(0)} - V^{(1)}\over
V^{(1)}}})}
\label{transf}
\end{eqnarray}
There exists two compatible Poisson brackets structures which endorse
our rational algebra of a bihamiltonian structure: the first Poisson
bracket structure, the original one, is determined by the
$sl(2)-{\cal KM}$ algebraic relations (\ref{KM}) for the fields
$J_{\pm,0}$; the second one is again determined from (\ref{KM}), but
after
taking into account the field transformations $J_-\mapsto J_-$, $J_+
\mapsto \cd J_+$. The compatibility
of the two Poisson brackets means that, for any function $f$ we have
the equality
\begin{eqnarray}
{\textstyle{d f\over d t}} &=& \{ H_1, f\}_2 = \{ H_2, f\}_1
\end{eqnarray}
The two Poisson brackets structures are the firsts of an infinite
series relating the infinite number of hamiltonians in involution.\\
The equations of motion relative to the first hamiltonian imply that
the  $V^{(n)}$ fields are free fields: ${\dot V^{(n)}} = {V^{(n)}}'$
for any $n$ (from now on we will use the standad conventions of dot
and prime to denote time and spatial derivative respectively).\\
The equations of motions for the fields $J_\pm$ relative to the
second hamiltonian are given by the following system (after setting
$J_0=0$):
\begin{eqnarray}
{\dot J_\pm} &=& \pm {J_\pm} '' \pm 2 J_\pm (J_+ J_-)
\label{nls}
\end{eqnarray}
This is nothing else than the coupled system associated to the NLS
equation; the standard NLS equation is recovered assuming the time
imaginary and setting
${{J_-}^{\star}}=J_+ = u$; such constraint is of course consistent
with the above equation. We get for $u$ the NLS equation in its
standard form:\cite{faddeev}
\begin{eqnarray}
i {\dot u} &=& u'' + 2 u|u|^2
\end{eqnarray}
In terms of the fields $V^{(n)}
$ the equations of motion relative to the second hamiltonian read as
follows:
\begin{eqnarray}
{\dot V^{(0)}} &=& 2\prt V^{(1)} - \prt^2 V^{(0)} \nonumber\\
{\dot V^{(1)}} &=& 2\prt V^{(2)} - \prt^2 V^{(1)}  + 2 V^{(0)}\prt
V^{(0)}
\label{eqmo}
\end{eqnarray}
These two equations, together with the algebraic constraints
(\ref{rela}) are sufficient to generate the whole tower of equations
of motion
\begin{eqnarray}
{\dot V^{(n)}} &=& 2\prt V^{(n+1)} - \prt^2 V^{(n)}
+ 2 \sum_{k=1}^n \left(\begin{array}{c} n\\ k\end{array}\right)
V^{(n-k)} \prt^k V^{(0)}
\label{eqmo2}
\end{eqnarray}
The NLS hierarchy is integrable, its integrability property being
made
manifest
from the existence of a Lax pair representation in terms of
pseudo-differential operators (PDO)(see \cite{dickey} for more
details and for the conventions here used).\\
Let $L =\prt + \sum_{n=0}^\infty u_n\prt^{-n-1}$ be the PDO of the KP
hierarchy. The different flows are defined through the position
\begin{eqnarray}
{\prt L\over \prt t_k}& = &[ L, {L^k}_+]
\label{lax}
\end{eqnarray}
where $k$ is a positive integer and ${L^k}_+$ denotes the purely
differential part of the operator.\\
The equations of motion of the NLS hierarchy relative to the second
hamiltonian
coincide with eq.(\ref{eqmo},\ref{eqmo2}) for $k=2$ after reduction,
namely after
constraining the infinite
number of fields $u_n$ to be $u_n(x,t_2) = (-1)^n V^{(n)}$.
Let us point out that, after having identified $u_0 \equiv V^{(0)}$,
$ u_1 \equiv - V^{(1)}$, the reduction of the remaining fields is
uniquely determined
by induction if we wish to reproduce from (\ref{lax}) the equations
of motion
(\ref{eqmo},\ref{eqmo2}): here again it is manifest the role of the
above
rational $\cw$ algebra as underlining structure which allows to
perform in a consistent way a reduction of the KP hierarchy having an
infinite set of
independent fields, to a KdV-like hierarchy involving only a finite
number of independent fields.

The Lax pair version of (\ref{eqmo},\ref{eqmo2}) ensures the
existence of an infinite number
of first integrals of motion $ K_r = < L^r > $, labelled by the
integers $r\geq 1$. Here, as commonly used, we have introduced the symbol $
<A> =\int dw a_{-1} (w) $ where $A$ is a generic pseudodifferential
operator $A = ...
+ a_{-1} \prt^{-1} +... $.\\
 Furthermore it can be explicitly checked that the first
integrals are all in involution (this can also be seen as a
consequence of the above mentioned bihamiltonian structure).\\
 The next hamiltonian of the hierarchy,
after $H_{1,2}$ is $H_3 = \int ( V^{(2)} + V^{(0)}V^{(0)} ) $. \\
The infinite tower of hamiltonians can in principle be computed with
an
algorithmic procedure.

In this section we have basically rederived well known results
concerning the NLS hierarchy. The important point however is that we
were able to do so having,
as only input, the existence of the rational algebra (\ref{coset}).
Our framework
can be generalized to produce other hierarchies from other rational
algebras. In the next section we will furnish another example of
that.

\section{A 3-field hierarchy coset construction}

\indent

In this section another example will be treated for convincing
that the relation between rational algebras and hierarchies is not
incidental.
It is based on the rational algebra introduced in subsection 2.2. It
is surely possible to derive the associated hierarchy directly from
the fractional KdV hierarchy of the Polyakov-Bershadsky
algebra:\cite{frac,genkdv} we plan to leave such derivation
for an extended version of this paper. Here we will repeat the steps
done in the previous section.

The basis of generating fields $T(w)$, $V^{(0,1)}$ for the rational
algebra
(\ref{commalgebra}) does not satisfy the property that the
coefficients of the
delta-function terms are all total derivatives. However, there exists
another basis, obtained from the previous one by simply replacing
\begin{eqnarray}
V^{(1)} \mapsto {\hat V^{(1)}} &=& V^{(1)} -
{\textstyle {1\over 4} } T^2
\end{eqnarray}
for which the property is satisfied.\\
It is clear that the two sets of fields are equivalent generating
sets for the rational algebra (\ref{commalgebra}). Notice that ${\hat
V^{(1)}}$ is, up to total derivative contributions, the unique
field of dimension $4$ (the dimension of both $V^{(1)}$ and $T^2$),
which satisfies the above property.\\
Therefore we have three hamiltonians, mutually in involution, which
are the first ones of the infinite hierarchy. They are
given by $H_1 = \int dw T(w)$, $H_2 =\int dw V^{(0)}(w)$ and $H_3
=\int dw {\hat V^{(1)}}(w)$. \\
As in the previous section, the dynamics for $T, V^{(n)}$ can be
reconstructed from the dynamics of $T, W_\pm$ and conversely, with
non-linear
transformations relating $W_\pm$ to $V^{(0,1)}$ as
eq.(\ref{transf}).

The equations of motion relative to the $H_1$ hamiltonian are the
free field equations: ${\dot T} = T' $, ${\dot V^{(n)}} =
{V^{(n)}}'$.\\
The system of equations of motion with respect to the $H_2$
hamiltonian
is the following one:
\begin{eqnarray}
{\dot T} &=& 2 {V^{(0)}}'\nonumber\\
{\dot W_\pm} &=& \pm {W_\pm}^{(2)} \pm T W_\pm
\label{system}
\end{eqnarray}
(here we have denoted ${W_\pm}^{(k)} \equiv \cd^k W_\pm$).\\
In terms of $V^{(0,1)}$ the above equations read:
\begin{eqnarray}
{\dot V^{(0)}} &=& 2 \prt V^{(1)} - \prt^2 V^{(0)}\nonumber\\
{\dot V^{(1)}} &=& 2 \prt V^{(2)} -\prt^2 V^{(1)} - V^{(0)}\prt T
\label{3feqmo}
\end{eqnarray}
For generic $V^{(n)}$, $n\geq 2$, the
equations are determined from the above ones and are explicitly given
by
\begin{eqnarray}
{\dot V^{(n)}} &=& 2 \prt V^{(n+1)} - \prt^2 V^{(n)}
-\sum_{k=1}^n \left( \begin{array}{c} n\\ k \end{array}\right)
V^{(n-k)} \prt^k T
\label{3feqmo2}
\end{eqnarray}
Here again the integrability properties are made manifest by the
existence of a Lax pair formalism. In this case the
pseudodifferential operator
$L$ is given by $L = \prt^2 +\sum_{n=0}^\infty u_n \prt^{-n}$. The
equations (\ref{3feqmo},\ref{3feqmo2}) are recovered from the flow
\begin{eqnarray}
{\prt L\over \prt t_1}& = &[ L, L_+]
\end{eqnarray}
once imposed the constraints $ u_0 (x, t_1) = T$ and  $u_{n+1}(x,t_1)
= (-1)^n V^{(n)}$ for $n\geq 0$.\\
The infinite tower of hamiltonians in involution can now be computed
with the usual standard procedure. The first three hamiltonians of
the hierarchy are the ones given above.

It is worth mentioning that the system (\ref{system}) coincides,
after field
redefinitions, with the hierarchy eq.(30) of ref.\cite{bonora}. The
latter being
a one-field reduction of the four-field representation of the KP
hierarchy.
It is indeed true that the above hierarchy admits three independent
fields.

It is tempting from eq.(\ref{system}) to set $T = 2 V^{(0)}$ and to
further reduce
the hierarchy (\ref{system}) to the NLS hierarchy studied in the
previous section.
Indeed the equations of motions for $W_\pm$ will look like eq.
(\ref{nls})
(after a change in sign is taken into account: $W_-\mapsto -W_-$).
This constraint plus the equations of motion make $T$ and $V^{(0)}$
free fields.
In order to be consistent however we have to require the
compatibility
of the dynamics of $T$
with the dynamics of $V^{(0)}$: this imply that $V^{(0)}$ must
satisfy
a constraint given by a differential equation; this constraint itself
must be compatible with the
dynamics, a further constraints is generated and so on. An infinite
tower of differential equations as constraints
on the linearly independent fields $V^{(n)}$ is produced;
the first two constraints of the series being given by
\begin{eqnarray}
&\prt (2 V^{(1)} - V^{(0)} -\prt V^{(0)} ) = 0&\nonumber\\
&\prt^2 ( 4 V^{(2)} - 4 \prt V^{(1)} -2 V^{(0)} V^{(0)} -2
V^{(1)}+\prt V^{(0)}
 +\prt^2 V^{(0)}) = 0. &\nonumber\\
&&
\end{eqnarray}
{}~\quad\\
\vfill
\newpage
{\Large {\bf Conclusions}}

\indent

In this letter we have emphasized the role played by rational $\cw$
algebras
in studying hierachies of integrable equations. In particular we have
shown
how to produce such hierachies from known rational $\cw$ algebras. A
natural
question arises: is the converse true? Any finite-field
representation of
KP hierarchies gives rise to an associated rational $\cw$ algebra? It
is
tempting to answer in an adfirmative way. If this would be the case,
then
rational $\cw$ algebras should be a fundamental tool for classifying
hierarchies of integrable equations. This problem being transferred
to the problem of classifing rational $\cw$ algebras. It is far from
being an easy
problem to solve\footnote{in the case of standard polynomial $\cw$
algebras,
all those arising as Toda reduction of a WZNW model are classified by
the non-equivalent $sl(2)$ embeddings in a given
Lie algebra \cite{sorba}. For a discussion on the completeness of
polynomial algebras so generated see \cite{classif}.}, but at least
is a well-posed problem (finite structures like rational algebras
seem more treatable objects than infinite ones, like non-linear
$\cw_\infty$ algebras).
In forthcoming papers we plan to address
these problems and to give at least a partial answer.\\ We should
also
mention here that
the quantum version of this work is in preparation. It is amazing
that rational algebras can be defined in the quantum case as well,
and work just like the corresponding classical versions.\\
 Last but not least we wish
to point out that having at disposal the explicit rational algebra
associated
to an integrable hierarchy is very illuminating and can help to
simplify proofs,
especially when concerned induction. We have already exploited such
property in
this letter.
{}~\\~\\

\noindent
{\large{\bf Acknowledgements}}
{}~\\~\\
I have profited of illuminating discussions had
with L. Bonora, F. Delduc, L. Feher,
L. Frappat, E. Ragoucy and, last but not least, P. Sorba.
{}~\\
{}~\\

\end{document}